\documentclass[aps,twocolumn,pra,showpacs,floatfix]{revtex4-1}

\usepackage{dcolumn}

\newcommand{\tm}{\tablenotemark}

\begin{document}

\title{\bf Spectra of barium, radium, and element 120; 
application of the combined correlation potential, singles-doubles, 
and configuration interaction {\it ab initio} method}
\author{J. S. M. Ginges}
\affiliation{School of Physics, University of New South Wales,
Sydney NSW 2052, Australia}
\author{V. A. Dzuba}
\affiliation{School of Physics, University of New South Wales,
Sydney NSW 2052, Australia}

\date{\today}

\begin{abstract}

We apply a version of the recently developed approach combining
the correlation potential, linearized singles-doubles
coupled-cluster, and the configuration interaction methods to the
spectra of the heavy alkaline earths barium, radium, and element 120. 
Quantum electrodynamics radiative corrections are included. We have found 
unprecedented agreement between {\it ab initio} theory and experiment 
for the spectra of barium and radium, and we make accurate predictions
for missing and unreliable data for all three atoms.

\end{abstract}


\maketitle

\section{Introduction}

Preparatory work with the radium atom towards measurements of fundamental symmetries violations 
is in progress at Argonne National Laboratory (USA) \cite{anl} and Kernfysisch
Versneller Instituut (The Netherlands) \cite{kvi}. 
Studies of the parity and time-reversal violating atomic electric dipole moment (EDM) and 
atomic parity violation (APV) are particularly attractive in radium
due to orders of magnitude enhancement of the effects, arising from
both nuclear and electronic mechanisms: 
the presence of octupole deformation of the nucleus may lead 
to several-hundred times nuclear enhancement of the EDM in radium in
the electronic ground state compared to mercury (see, e.g., \cite{afs_schiff,de_schiff}), 
for which the best limit on an atomic EDM has been placed \cite{hg};
the presence of anomalously close electronic levels of opposite
parity may lead to orders of magnitude electronic enhancement of
EDM effects in metastable states and APV effects in certain
transitions \cite{flambaum_Ra,radiumDFG}.

High-precision atomic structure calculations will be required for
interpretation of the APV and atomic EDM measurements \cite{ginges_review}. Moreover, the
measured low-lying excitation energies for radium are incomplete, 
and it is important to have high-precision predictions for these
levels.
There are already a number of theoretical works devoted to
studies of fundamental symmetries violations in radium 
\cite{radiumDFG,DFGK2002,BGGFIJ,latha2013,radiumRDF,RGJB2014,latha2014}
and to the radium spectra, lifetimes, and hyperfine structure 
\cite{landau,BFFP2004,BP2005,bieron2005,dzuba_ginges_Ra,dzuba_flambaum_Ra,BIJ2007}.
There are also recent measurements of a few transition frequencies and
lifetimes \cite{scielzo2006,trimble2009,santra2014}.
In this work we use a recently-developed approach that combines 
the correlation potential, singles-doubles coupled-cluster,
and configuration interaction methods \cite{dzuba_sdci} to the radium
spectra. We have found unprecedented agreement between theory and
experiment for most levels using this {\it ab initio} method. 

The heavier electronic homologue of radium is element 120. Efforts to
synthesize this element are underway at GSI (Germany), JINR (Russia),
and RIKEN (Japan) (see, e.g., \cite{120_synth}). This element lies within the predicted
island of stability, a region of increased stability against nuclear
decay close to the next doubly magic shell closures Z=114, 120, or
126 and N=172 or 184, depending on the model \cite{magic}.
If Z=120 is the next closed proton shell, significantly increased stability is
expected for this element. Enhanced stability could make atomic and chemical studies
of this element possible. Already, chemical studies involving the
superheavy element Sb (Z=106) have proved successful \cite{Sb_chemistry}.  
The current calculations for E120 extend the spectral range considered in 
previous works \cite{dinh_E120,skripnikov,borschevsky}.

We also perform calculations for the lighter electronic homologue
barium, as this can be used to gauge the accuracy for the heavier 
elements. The method used in this work has already been applied 
to the low-lying levels of barium in Ref. \cite{dzuba_sdci}. Here we extend the application to
the excitation energies of the lowest 31 levels. 

\section{Method of calculation}

We use an approach that is based on the combination of several different many-body methods: 
the correlation potential (CP), linearized singles-doubles coupled-cluster (SD), and configuration 
interaction (CI) methods. It may be referred to as the CP+SD+CI method. This method was recently 
developed by Dzuba in Ref. \cite{dzuba_sdci} and is similar to
the combined SD+CI method developed by Safronova {\it et al}. \cite{safronova_sdci}. 

The CP+SD+CI and SD+CI methods are essentially based on the method
combining many-body perturbation theory and the configuration interaction (MBPT+CI) \cite{dfk}, which has proven to be 
one of the most computationally efficient and accurate approaches for calculations involving heavy atoms with 
several valence electrons. These methods differ in their treatment of
the valence-core correlations and the screened Coulomb interaction
(valence-core-valence correlations), as we explain below. 

Barium, radium, and element 120 have two valence electrons above a closed electronic core. 
We wish to find the eigenvalues and eigenvectors of the effective
Hamiltonian for the two valence electrons, 
\begin{equation}
H^{CI}=h_1(r_1)+h_1(r_2)+h_2(r_1,r_2) \ ,
\end{equation}
by diagonalizing $H^{CI}$ with respect
to wavefunctions constructed from linear combinations of two-electron
Slater determinants; the Slater determinants are formed from
relativistic Hartree-Fock orbitals found in the core of $N-2$ electrons.
Here $h_1$ contains all one-electron terms of the 
many-electron problem and $h_2$ contains all two-electron terms.
The one-electron terms are
\begin{equation}
h_1=c \mbox{\boldmath{$\alpha$}}\cdot {\bf p}+(\beta -1)mc^2 + V_{\rm nuc}+ V_{HF}^{(N-2)}+\Sigma_1 \ ,
\end{equation}
where $\mbox{\boldmath{$\alpha$}}$ and $\beta$ are Dirac matrices,
$V_{\rm nuc}$ is the nuclear potential (we use the two-parameter Fermi
distribution for the nuclear density), 
$V_{HF} ^{(N-2)}$ is the self-consistent Hartree-Fock potential of the $N-2$ electrons making up 
the electronic core, and 
$\Sigma _1$ is the correlation potential which accounts for the correlations between a single valence 
electron and the core. 
For E120, the choice of the nuclear parameters is important; for our
Fermi distribution, we use a nuclear radius of $r=8.0\, {\rm fm}$ and
nuclear thickness $t=2.0\, {\rm fm}$, corresponding to a
root-mean-square radius $r_{\rm rms}\approx 6.42 \, {\rm fm}$. These are the parameters we used
in our previous calculation \cite{dinh_E120}, and we make this choice again
for easy comparison with that work. 

The two-electron terms $h_2$ consist of the Coulomb interaction and 
the two-electron correlation operator $\Sigma_2$ (the screened
Coulomb interaction),
\begin{equation}
h_2(r_1,r_2) = \frac{e^2}{|r_1-r_2|} + \Sigma_2(r_1,r_2) \ .
\end{equation} 

In the MBPT+CI method, both
$\Sigma_1$ and $\Sigma_2$  are calculated using MBPT;  
they are often calculated in only second-order in the Coulomb
interaction. We have demonstrated before, that using the all-order correlation potential 
$\Sigma ^{\infty}_1$ in place of the second-order correlation potential 
$\Sigma ^{(2)}_1$ leads to significantly improved accuracy
\cite{dzuba_ginges_Ra}. We refer to this latter approach, with the all-order
$\Sigma^{\infty}_1$, as the CP+CI method. In the SD+CI method \cite{safronova_sdci}, both 
$\Sigma_1$ and $\Sigma_2$ are found from all-order linearized
singles-doubles coupled-cluster equations. 

In the current 
CP+SD+CI approach, 
we calculate the all-order $\Sigma_1^{\infty}$ 
using the Feynman diagram technique, 
while the screening of the Coulomb interaction $\Sigma_2$ is
found from the SD method. In this way, we include important 
classes of many-body diagrams to all-orders in the Coulomb
interaction, both for $\Sigma_1$ and $\Sigma_2$. 
It was shown in Ref. \cite{dzuba_sdci} that high accuracy in
excitation energies can be obtained with this choice.

\begin{table}
\caption{Factors for Ba used to mimic higher-order screening for
  exchange diagrams, $f_k=\langle v|\Sigma^{\infty,ee}_{{\rm dir},k}|v \rangle / \langle v
  |\Sigma^{(2)}_{{\rm dir},k}|v\rangle$ and $f_k^{\rm hp}=\langle
  v|\Sigma^{\infty}_{{\rm dir},k}|v \rangle / \langle v
  |\Sigma^{(2)}_{{\rm dir},k}|v\rangle$; $k$ is the multipolarity of the Coulomb interaction.}
\label{tab:screening}
\begin{ruledtabular}
\begin{tabular}{lcccccccccc}
$k$ & 0 & 1& 2& 3& 4& 5& 6 & 7 & 8 & 9 \\ 
\hline
$f_k$ & 0.73 & 0.61 & 0.83 & 0.91 & 0.99 & 1.05 & 1.10 & 1.00 & 1.00 & 1.00 \\
$f_k^{\rm hp}$ &0.86 & 0.73 & 0.97 & 1.02 & 1.07 & 1.11 & 1.15 & 1.00 & 1.00 & 1.00 \\
\end{tabular}
\end{ruledtabular}
\end{table}

In the calculation of $\Sigma_1$, the Feynman diagram technique is
used to include two classes of diagrams to all orders in the Coulomb interaction for 
the {\it direct} part: 
the electron-electron Coulomb screening and the hole-particle
interaction inside the internal loops \cite{corr_pot}. The {\it exchange} part is evaluated  
in second-order in the Coulomb interaction by calculation of the 
corresponding Goldstone diagrams. Screening is 
taken into account in a simplified way, by multiplying the Coulomb integrals by 
factors found from the direct diagrams (more on this below). 
There is another series of diagrams, referred to as ``ladder
diagrams'', that are calculated using singles-doubles-type 
equations \cite{ladder};
the corresponding corrections to the valence energies are added to the 
CI matrix.

Screening factors for the exchange diagrams are found by taking the ratio of the 
expectation value for the direct part of $\Sigma_1^{\infty,ee}$ (with the dominant all-orders
electron-electron screening considered) to the expectation value for the
direct part of $\Sigma_1^{(2)}$ for each multipolarity $k$, that is,
 $f_k=\langle v|\Sigma^{\infty,ee}_{{\rm dir},k}|v \rangle / \langle v
 |\Sigma^{(2)}_{{\rm dir},k}|v\rangle$.
The corresponding factors for Ba 
are shown in the first row of Table \ref{tab:screening}.
Later, as a test of our accuracy, we consider the effect of using a 
different set of screening factors $f_k^{\rm hp}$, found by considering both the
dominant electron-electron screening and the hole-particle interaction 
series of diagrams; these factors are listed in the second row of Table
\ref{tab:screening}.   

Note that we used slightly different factors for $f_k$ in different
approximations (with or without Breit or QED corrections) as well as
slightly different factors for different atoms (Ba, Ra, E120). They 
differ by a few percent at most, and as their precise values are not
of general interest, we do not present them here.

We refer the reader to Ref. \cite{dzuba_sdci} for details regarding the 
calculation of $\Sigma_2$. 

In this work we quantify the corrections associated with inclusion of 
the Breit interaction and the quantum electrodynamics radiative
corrections. The Breit interaction is considered at the relativistic
Hartree-Fock (RHF) level. The 
radiative corrections are taken into account through the addition of a local radiative potential
\cite{radpot} to the nuclear potential. This radiative potential method provided a breakthough in the ability for radiative 
corrections to be included into the many-body problem in heavy atoms. 
We refer the reader to Ref. \cite{radpot} for details about this method. The radiative
potential method has since been implemented in a number of works, including
(with a minor modification to one of the fitting factors in the radiative potential)
Refs. \cite{thierfelder,HDS2013}, where the radiative potential was added to the atomic many-body
package GRASP \cite{grasp1989,grasp2006}.

The upper and lower radial components of the relativistic Hartree-Fock
orbitals in the $V^{N-2}$ potential 
are expanded in a basis of B-splines of order $k=9$ \cite{johnson}. 
We use 40 B-splines for Ba and Ra and 50 B-splines for E120 confined to a cavity of radius
40 a.u. 
We perform the calculations for the exchange
part of the correlation potential, the ladder diagrams, and $\Sigma_2$ using the lowest 30 states 
in each partial wave up to $l=6$ for the intermediate excitations. 
We include the correlation potential $\Sigma_1$ for the valence levels up
to $l=4$ and $\Sigma_2$ for the valence levels 
up to $l=3$.
At the CI stage of the calculations, the basis for Ba and Ra (and E120)
consists of the lowest 14 (18) valence states in each partial wave up to $l=4$.

To quantify the size of the Breit and QED corrections, we have
carried out three runs of the calculations, for both the ions and the
neutral divalent atoms: with neither Breit nor QED, with Breit and without QED,
and with both Breit and QED. As the basis is determined by the
Hartree-Fock orbitals, and Breit and QED
corrections enter at this level, we have used a different basis
set for each of these runs. For each run, one basis is used for 
{\it all} aspects of the many-body problem: CP, SD, and CI.

\section{Results}

Our calculations begin for the ions Ba$^+$, Ra$^+$, and E120$^+$. The
correlation potential $\Sigma_1$ that describes the valence-core 
correlations is the same  for the monovalent ions above as for the 
divalent atoms Ba, Ra, E120 when we perform the RHF calculations in 
the field of the $N-2$ electrons of the core. Therefore, the quality of the spectra 
for the ions is a good indication of the quality of the correlation 
potential $\Sigma_1$ \cite{dzuba_ginges_Ra}. 

\subsection{Ions}

\label{ssec:ions}

Results for the ionization energies of the
lowest partial waves up to $l=3$ for the ions Ba$^+$, Ra$^+$, and 
E120$^+$ are presented in Tables \ref{tab:Ba_ion}, \ref{tab:Ra_ion},
and \ref{tab:E120_ion}, respectively. 
We separate the contributions arising from inclusion of the
correlation potential $\Sigma_1^{\infty}$ (without ladder diagrams),
ladder diagrams ``Lad.'', the 
Breit interaction, and QED radiative corrections.
 
Absolute differences of our final results from
experiment are listed in the last column under
``$\Delta$''. These differences for Ba$^+$ and Ra$^+$ are very small,
on the order of $10-100\, {\rm cm}^{-1}$,
with the largest difference $|\Delta| \approx 160\, {\rm cm}^{-1}$. We expect a
similar level of agreement for E120$^+$.

\begin{table}
\caption{Ionization energies for the lowest states in each wave to
  $l=3$ for Ba$^{+}$; units ${\rm cm}^{-1}$.}
\label{tab:Ba_ion}
\begin{ruledtabular}
\begin{tabular}{lrrrrrrrr}
State & RHF & + $\Sigma_1^{\infty}$ & Lad. & Breit & QED & Total &
Exp.\tablenotemark[1] & $\Delta$\\
\hline
6$s_{1/2}$               & 75340 & 80808 & -156 &-22 &-45 & 80585&80686 & -101 \\ 
 5$d_{3/2}$       & 68139 & 76466 & -763 &34&36 & 75773 &75812&-39\\
      5$d_{5/2}$       & 67665 & 75622 & -765 & 57&32 & 74946 &75011& -65 \\
      6$p_{1/2}$        & 57266 & 60591 & -128 &-36 &5 & 60432 &60425& 7 \\
      6$p_{3/2}$       & 55873 & 58865 & -118 & -16 &3 & 58734 &58734 & 0\\
     4$f_{5/2}$        & 28213 & 32180 & -32 &117 &32 & 32297 &32428 & -131\\
     4$f_{7/2}$       & 28222 & 31989 & -32 &126 &28  & 32112 &32203& -91\\
\end{tabular}
\end{ruledtabular}
\tablenotetext[1]{NIST data, Ref.~\cite{NIST}.}
\end{table}

We can see from a comparison of the final results ``Total'' in Tables \ref{tab:Ba_ion},
\ref{tab:Ra_ion}, \ref{tab:E120_ion} the effect of the 
relativistic contraction of the $s$ and $p_{1/2}$ orbitals in the
heavier homologues, which in turn screen and push out the 
orbitals with higher angular momenta.
 
Inclusion of the ladder diagrams is very important
for reaching good agreement for the $d$ levels. 
The $d$-orbitals are very low-lying in the ions and play a big role in the
low-lying spectra of the neutral divalent atoms. The size of the
ladder diagram contribution decreases as we go from Ba$^+$ to E120$^+$, as the $d$
orbitals are pushed out, while the contributions for $s$ and $p$ levels increase.

The contributions from Breit and QED are roughly of the
same magnitude, and generally increase with higher $Z$. The QED
corrections exceed $100\, {\rm cm}^{-1}$ for the $s$ levels for
E120$^{+}$, while they are negligible for the $p$ waves. 
The largest Breit corrections we see are for the
$f$ levels, almost reaching $200\, {\rm cm}^{-1}$. Interestingly, 
these $f$-wave corrections are mostly determined by many-body effects 
through the inclusion of $\Sigma_1^{\infty}$, and they are sensitive
to the choice of the correlation potential.

\begin{table}[h]
\caption{Ionization energies for the lowest states in each wave to
  $l=3$ for Ra$^{+}$; units ${\rm cm}^{-1}$.}
\label{tab:Ra_ion}
\begin{ruledtabular}
\begin{tabular}{lrrrrrrrr}
State & RHF & + $\Sigma_1^{\infty}$ & Lad. & Breit & QED & Total &
Exp.\tablenotemark[1] & $\Delta$\\
\hline
      7$s_{1/2}$               & 75898 &82010&-219 &-21 &-87&81684 &  81842 &-158\\
        6$d_{3/2}$       & 62356 &70242&-620& 46 &38& 69707&  69758 & -51\\
        6$d_{5/2}$       & 61592 &68518 &-643&67&31& 67973 & 68099 & -126\\
       7$p_{1/2}$        & 56878 &60739&-182&-59&2& 60499 & 60491 & 8\\
        7$p_{3/2}$       & 52906 &55771&-140&-18&-2& 55611 & 55633 & -22\\
       5$f_{5/2}$        & 28660 &32768& -74 &109&32& 32835 & 32854 & -19\\
        5$f_{7/2}$       & 28705 &32542& -76 &107&27& 32600 & 32570 & 30\\
\end{tabular}
\end{ruledtabular}
\tablenotetext[1]{NIST data, Ref.~\cite{NIST}.}
\end{table}

It is worthwhile
pointing out that the radiative potential \cite{radpot} was found by
fitting to the $s$ and $p$ levels for the hydrogen-like ions; it was 
not fitted to higher waves. The radiative QED corrections for the
$d$ and $f$ waves, however, largely arise due to account of many-body effects, 
namely core relaxation and the correlation potential. 
Core relaxation corresponds to alteration of the electronic core due 
to self-consistent solution of the RHF equations with the radiative
potential included. 
The largest part of the core relaxation correction to the valence level shift occurs as a result of radiative 
corrections to $s$-orbitals in the core, which are well-fitted. 

We have previously calculated the spectra of these ions for $s$ and
$p$ levels, including estimates of Breit and radiative corrections \cite{dinh_E120_ions}; more recently the 
spectra was calculated in Ref. \cite{dzuba_ions} for the lowest $s$, $p$, and $d$
levels, this time with the contribution from 
ladder diagrams, also.

\begin{table}[h]
\caption{Ionization energies for the lowest states in each wave to
  $l=3$ for E120$^{+}$; units ${\rm cm}^{-1}$.}
\label{tab:E120_ion}
\begin{ruledtabular}
\begin{tabular}{lrrrrrrrr}
 State & RHF & + $\Sigma_1^{\infty}$ & Lad. & Breit & QED & Total \\
\hline
       8$s_{1/2}$             & 83168 &90105&-520&-108&-129& 89349\\
       8$p_{1/2}$      & 60027 &65475&-379 &-141&-9& 64946\\
       7$d_{3/2}$      & 56620 &64841& -593&51&58& 64357\\
       7$d_{5/2}$      & 56413 &62684& -625 &46&46&62151\\
       8$p_{3/2}$       & 49295 &52017&-179 &-18&-7& 51812\\
       6$f_{5/2}$       & 29734 &36907&-211 &186&97&36978\\
      6$f_{7/2}$      &29909&36252&-219 &169&77&36279\\
\end{tabular}
\end{ruledtabular}
\end{table}

\subsection{Divalent atoms}

In Tables \ref{tab:Ba}, \ref{tab:Ra}, and \ref{tab:E120} we present
our results for the ionization potentials (removal of one $s$ electron,
IP1, and removal of both $s$ electrons, IP1+IP2) 
and excitation energies for the lowest 31 levels for Ba, lowest 40
levels for Ra, and lowest 25 levels for E120. Results presented under 
the column ``CP$^{\rm nl}$+SD+CI" mean that 
the all-orders correlation potential $\Sigma^{\infty}_1$ is included 
(though ladder diagrams are not taken into account; the superscript ``nl'' is
short for ``no ladder'') and that the all-orders $\Sigma_2$ is 
included (calculated using the SD method).
Contributions from ladder diagrams, Breit, and QED radiative corrections 
appear in the following columns. 

Our final results are presented under the column ``Total'', and the 
column ``$\Delta_{\rm Exp}$'' gives the deviation of these results from
experiment, $\Delta_{\rm Exp} = {\rm Total} - {\rm Exp}$. Results of
other calculations are given in the tables, also.
In particular, we
have presented the most precise {\it ab initio} calculations
available. These include different versions of
coupled-cluster \cite{landau,mani_angom,skripnikov,borschevsky},
CP+CI \cite{dinh_E120}, and SD+CI \cite{safronova_sdci}. 
For the higher levels, where there is only limited data available, 
we also present for Ba and Ra the results of semi-empirical CP+CI
calculations \cite{dzuba_flambaum_Ra}.

\subsubsection{Barium}

\begin{table*}
\caption{Ionization potentials (removal of one electron, IP1, and both
  electrons, IP1+IP2) and excitation energies for Ba. 
CP$^{\rm nl}$+SD+CI excludes ladder diagrams, Breit, and QED; 
these corrections are given in subsequent columns. 
$\Delta_{\rm Exp}={\rm Total}-{\rm Exp}$.
Results of other calculations are presented in the final
column. Units: ${\rm cm}^{-1}$.}
\label{tab:Ba}
\begin{ruledtabular}
\begin{tabular}{lllrrrrrrrl}
Conf. &  Term & J & CP$^{\rm nl}$ &Ladder& Breit & QED & Total
& Exp.\tablenotemark[1] & $\Delta_{\rm Exp}$ & Other \\
 & & & +SD+CI & & & & & &&\\
\hline
 IP1\qquad \ \ $6s^2$ &  ${^1S}$&0&42097&-173&122&-19&42027&42035&
 -8& 42444\tm[2], 42120\tm[3]\\
 IP1+IP2\ $6s^2$ &  ${^1S}$&0&122905&-329&100&-64&122612&122721&
 -109&122954\tm[3], 123363\tm[4], 123869\tm[5]\\
     $6s5d$ & ${^3D}$ &1&8460&575&-70&-60&8905&9034&-129&9117\tm[2], 8687\tm[3], 9249\tm[4], 9077\tm[5]\\
                 &&2&8704&576&-68&-59&9153&9216&-63&9296\tm[2], 8875\tm[3], 9441\tm[4], 9369\tm[5]\\
                 &                 &3&9066&576&-61&-57&9524&9597&-73&9677\tm[2], 9279\tm[3], 9840\tm[4], 9830\tm[5]\\
     $6s5d$ & ${^1D}$ &2&10996&589&-62&-66&11457&11395&62&11426\tm[2], 11081\tm[3], 11727\tm[4], 11871\tm[5]\\
     $6s6p$ & ${^3P}^{\rm o}$ &0&12351&-33&27&-35&12310&12266&44&12357\tm[2],12099\tm[3], 12556\tm[4], 12668\tm[5]\\
                  &   &1&12706&-33&26&-35&12664&12637&27&12728\tm[2], 12474\tm[3], 12919\tm[4], 12947\tm[5]\\
                  & &2&13603&-41&25&-32&13555&13515&40&13610\tm[2], 13365\tm[3], 13819\tm[4], 13449\tm[5]\\
     $6s6p$ & ${^1P}^{\rm o}$ &1&17887&107&11&-41&17964&18060&-96&18170\tm[2], 17943\tm[3], 18292\tm[4], 20077\tm[5]\\
     $5d^2$ & ${^3F}$ &2&19920&1142&-119&-110&20833&20934&-101&21017\tm[2]\\
                 &              &3&20142&1146&-115&-108&21065&21250&-185&21338\tm[2]\\
                 &              &4&20636&1145&-109&-106&21566&21624&-58&21714\tm[2]\\
     $5d6p$ & ${^3F}^{\rm o}$ &2&21545&545&-46&-96&21948&22065&-117&22238\tm[2]\\
                  &                 &3&22436&538&-42&-93&22839&22947&-108&23116\tm[2]\\
                  &                  &4&23286&534&-39&-91&23690&23757&-67&23950\tm[2]\\
     $5d^2$ & ${^1D}$ &2&22182&996&-86&-103&22989&23062&-73&23077\tm[2]\\
     $5d6p$ & ${^1D}^{\rm o}$ &2&22607&543&-48&-94&23008&23074&-66&23289\tm[2]\\
     $5d^2$ & ${^3P}$     &0&21985&1001&-88&-104&22794&23209&-415&23213\tm[2]\\
                 &                  &1&22118&1022&-90&-104&22946&23480&-534&23500\tm[2]\\
                &                   &2&22897&1003&-81&-100&23719&23919&-200&23950\tm[2]\\
     $5d6p$ & ${^3D}^{\rm o}$ &1&23635&541&-61&-98&24017&24192&-175&24474\tm[2]\\
                  &                 &2&23992&545&-58&-97&24382&24532&-150&24817\tm[2]\\
                  &                 &3&24460&543&-56&-95&24852&24980&-128&25264\tm[2]\\
     $5d^2$ & ${^1G}$     &4&24328&1069&-89&-103&25205&&&24300(300)\tm[6], 24684\tm[2]\\ 
     $5d6p$ & ${^3P}^{\rm o}$  &0&25115&507&-54&-94&25474&25642&-168&25886\tm[2]\\
                  &                  &1&25183&503&-51&-92&25543&25704&-161&25947\tm[2]\\
                  &                  &2&25466&512&-46&-90&25842&25957&-115&26203\tm[2]\\
     $6s7s$ & ${^3S}$ &1&26253&-135&28&-15&26131&26160&-29&26573\tm[2]\\
     $5d^2$ & ${^1S}$ &0&25144&612&-28&-70&25658&26757&-1099&26034\tm[7]\\
     $5d6p$ & ${^1F}^{\rm o}$ &3&26343&498&-56&-94&26691&26816&-125&26968\tm[2]\\
    $6s7s$ & ${^1S}$ &0&28192&55&10&-37&28220&28230&-10&28583\tm[2]\\
     $5d6p$ & ${^1P}^{\rm o}$ &1&28351&210&-15&-59&28487&28554&-67&28788\tm[2]\\
\end{tabular}
\end{ruledtabular}
\tablenotetext[1]{NIST data, Ref.~\cite{NIST}.}
\tablenotetext[2]{IHFSCC, Ref.~\cite{landau}.} 
\tablenotetext[3]{CP+CI, Ref.~\cite{dinh_E120}. IP1 found in
  combination with the calculation for the ion, Ref.~\cite{dinh_E120_ions}.}
\tablenotetext[4]{SD+CI, Ref.~\cite{safronova_sdci}.}
\tablenotetext[5]{FSCC, Ref.~\cite{mani_angom}.}
\tablenotetext[6]{Estimated in the experimental work of Palenius, Ref.~\cite{palenius}.}
\tablenotetext[7]{Semi-empirical CP+CI, Ref.~\cite{dzuba_flambaum_Ra}.}
\end{table*}

For Ba, it is seen that the ladder diagrams give a contribution to the
excitation energies of about $500\, {\rm cm}^{-1}$ for configurations 
containing a single $5d$ orbital, that is for $6s5d$ and $5d6p$ (excluding 
the very highest level). For the $5d^2$ configuration, the
contribution amounts to roughly double this, $\approx 1000\, {\rm
  cm}^{-1}$. The ladder diagrams give a small correction to the other levels.

The Breit and radiative corrections for Ba are roughly of the same magnitude, 
ranging between about $10-120\, {\rm cm}^{-1}$ for Breit and $15-110\, {\rm
  cm}^{-1}$ for the radiative corrections for the levels
considered. For the higher levels the radiative corrections dominate.

We see that the deviation of our final results ``Total'' from
experiment, $\Delta_{\rm Exp}$, ranges between 
$10-200\, {\rm cm}^{-1}$, with the exception of the 
larger deviation for $5d^2\, ^3P_{0,1}$ of about $400\, {\rm
  cm}^{-1}$ and $500\, {\rm cm}^{-1}$ and the very large deviation 
for $5d^2\, ^1S_0$ of about  $1100 \, {\rm cm}^{-1}$.

The singlet state $5d^2~^1G_4$ has not been measured and is absent 
in the NIST data \cite{NIST}. Its 
position was predicted in the early experimental work of Palenius
\cite{palenius} to be $24300\pm 300~{\rm cm}^{-1}$. It has
subsequently been calculated in the works
\cite{RPG78,EKI96,DJ98,landau}; there are other calculations of the
barium spectra where this level has been missed.  
We present in the table, alongside our own result, the initial
estimate \cite{palenius} and the value from the most precise calculations 
\cite{landau}.   
Our calculations give the value $25205\, {\rm cm}^{-1}$ for this
level; looking at the results for other terms with the configuration
$5d^2$, 
we expect that the result for this level could be 
underestimated, possibly by as much as $100-300\, {\rm cm}^{-1}$.

Our result for $5d^2\, ^1S_0$ disagrees with the experimental value by 
$\approx 1100\, {\rm cm}^{-1}$. This very large difference is well
outside the deviations we see for the other levels.
This level does not
appear in the extensive spectra calculations of Landau {\it et al.}
\cite{landau}. We know of only one other calculation of this
level, carried out in the CP+CI method with empirical fitting
\cite{dzuba_flambaum_Ra}, with the result $26034\,{\rm cm}^{-1}$; this 
is about $700\,{\rm cm}^{-1}$ less than experiment, and well above 
the estimated error in that work. 

We should note, however, that the largest deviations we see are for
terms belonging to 
$5d^2$, and the unmeasured level $5d^2\, ^1G_4$ and
the level $5d^2\, ^1S_0$ belong to this configuration. It is possible
that our calculations do not do an adequate job for these low-lying
$d^2$ states. We note further the large difference between our result
and that of Ref. \cite{landau} for the energy of the level $5d^2\, ^1G_4$ .

The results of the current work are in significantly better agreement
with experiment over the considered spectral range compared to the {\it ab initio} calculations performed using CP+CI
\cite{dinh_E120}, SD+CI \cite{safronova_sdci}, and a version of Fock-space
coupled-cluster (FSCC) \cite{mani_angom}.
There is only one other work where comparable accuracy was reached,
using the intermediate Hamiltonian Fock-space coupled-cluster (IHFSCC) method
\cite{landau}; however, in that method the accuracy deteriorates for the higher levels. 

We should note that neither Breit nor QED corrections were included in
the results of Ref. \cite{dinh_E120}, while QED corrections were not
included in those of 
Refs. \cite{landau,safronova_sdci,mani_angom}. The QED correction to
IP1 was calculated in the work \cite{thierfelder} using the radiative
potential \cite{radpot}, and the value $-19\, {\rm cm}^{-1}$ was obtained -  the same 
result we have obtained in this work.  

\subsubsection{Radium}

\begin{table*}
\caption{Ionization potentials (removal of one electron, IP1, and both
  electrons, IP1+IP2) and excitation energies for Ra. 
CP$^{\rm nl}$+SD+CI excludes ladder diagrams, Breit, and QED; 
these corrections are given in subsequent columns. 
$\Delta_{\rm Exp}={\rm Total}-{\rm Exp}$.
Results of other calculations are presented in the final
column. Units: ${\rm cm}^{-1}$. Asterisks (*) identify where 
configurations have been modified from those in the NIST data
\cite{NIST} or where calculated energies in the final column have 
been reassigned to different terms.}
\label{tab:Ra}
\begin{ruledtabular}
\begin{tabular}{llrrrrrllrl}
Conf.& Term & J &CP$^{\rm nl}$&Ladder&Breit&QED& Total &
Exp.\tablenotemark[1] &$\Delta_{\rm Exp}$ & Other \\
 & & &+SD+CI &&& &&& & \\
\hline
IP1\qquad \ \ $7s^2$&
${^1S}$&0&42680&-227&271&-45&42679&42573&102&42622\tm[2], 42562\tm[3]\\ 
IP1+IP2\ $7s^2$&  ${^1S}$&0&124690 &-446 &250&-132&124363&124416&-53&124656\tm[2], 124642\tm[3]\\
 $7s7p$ & ${^3P}^{\rm o}$ &0&13173&-43&70&-64&13136&13078&58&12916\tm[2], 13093\tm[3]\\
         &  &1&14080&-47&71&-64&14040&13999&41&13844\tm[2], 14017\tm[3]\\
                  &                  &2&16828&-84&78&-60&16762&16689&73&16566\tm[2], 16675\tm[3]\\
     $7s6d$ & ${^3D}$     &1&13411&380&-46&-92&13653&13716&-63&13622\tm[2], 14021\tm[3]\\
                  &                  &2&13771&384&-39&-91&14025&13994&31&13902\tm[2], 14292\tm[3]\\
                  &                  &3&14440&401&-20&-85&14736&14707&29&14645\tm[2], 14989\tm[3]\\
     $7s6d$ & ${^1D}$     &2&16996&385&-12&-97&17272&17081&191&17004\tm[2], 17376\tm[3]\\
     $7s7p$ & ${^1P}^{\rm o}$     &1&20632&-34&69&-61&20606&20716&-110&20667\tm[2], 20792\tm[3]\\
     $7s8s$ & ${^3S}$     &1&26887&-181&73&-37&26742&26754&-12&26665\tm[4], 26762\tm[3]\\
     $7s8s$ & ${^1S}$ &0&27910&-119&70&-45&27816& &&27768\tm[4], 28248\tm[5]\\
     $6d7p$ & ${^3F}^{\rm o}$     &2&27803&337&26&-161&28005&28038&-33&27991\tm[4], 28328\tm[3]\\
                  &                 &3&29924&328&46&-155&30143&30118&25&30067\tm[4], 30388\tm[3]\\
                  &                  &4&32248&303&59&-151&32459&32368&91&32363\tm[4], 32603\tm[3]\\
      $6d^2$ &   ${^3F}$      &2&28964&738&-67&-180&29455&  &&29731\tm[4], 29610\tm[5]\\
                   &           &3&29648&776&-56&-179&30189&        &&30464\tm[4], 30404\tm[5]\\
                   &           &4&30455&787&-39&-175&31028&       &&31172\tm[4], 31114\tm[5]\\
     $6d^2$* &  $^3P$   &0&29426&402&32&-155&29705& &&29840\tm[4], 29833\tm[5]\\
                   &              &1&30659&560&10&-164&31065&31249&-184&31365\tm[4], 31342\tm[5]\\
                  &           &2&31892&1257&22&-142&33029&32941&88&33180\tm[4], 33147\tm[5]\\
     $7s8p$  &  $^1P^{\rm o}$ &1&30691&51&55&-94&30703&32858&-2155&30695\tm[4]*\\
     $6d^2$ & $^1D$ &2&30490&493&13&-144&30852& &&30982\tm[4], 30930\tm[5]\\
     $6d7p$ & ${^1D}^{\rm o}$ &2&30747&312&43&-150&30952&30918&34&30894\tm[4], 31178\tm[5]\\
     $7s8p$ & ${^3P}^{\rm o}$     &0&31180&-152&79&-53&31054&31086&-32&31008\tm[4], 31126\tm[3]\\
                 &     &1&31511&-5&55&-88&31473&31563&-90&31446\tm[4]*, 31636\tm[3]\\
                 &  &2&31913&-92&75&-64&31832&31874&-42&31778\tm[4], 31934\tm[3]\\
       $7s7d$ &   ${^3D}$       &1&32079&-144&74&-51&31958&32001&-43&31895\tm[4], 32423\tm[5]\\
                    &                   &2&32749&-810&69&-65&31943&31993&-50&31902\tm[4]\\
                  &                  &3&32273&-143&77&-51&32156&32197&-41&32068\tm[4], 32625\tm[5]\\
    $7s7d$*  &  ${^1D}$  &2&32171&26&53&-84&32166&32215&-49&32205\tm[4],32564\tm[5]\\
     $6d7p$  & ${^3D}^{\rm o}$     &1&32180&-1&52&55&32136&32230&-94&32090\tm[4], 32614\tm[5]\\
                   &                   &2&32287&279&17&-157&32426&32507&-81&32436\tm[4], 32846\tm[5]\\
                   &                   &3&33021&254&28&-147&33156&33197&-41&33169\tm[4], 33531\tm[5]\\
     $6d7p$   & ${^3P}^{\rm o}$     &0&33543&247&21&-151&33660&33782&-122&33809\tm[4], 34055\tm[5]\\
                    &     &1&33586&239&24&-147&33702&33824&-122&33837\tm[4], 34102\tm[5]\\
                    &                    &2&34193&283&43&-147&34372&34383&-11&34421\tm[4], 34677\tm[5]\\
     $6d^2$   &   ${^1G}$   &4&33142&724&-16&-166&33684& &&33261\tm[5]\\
     $6d^2$   &  ${^1S}$                    &0&33561&312&54&-116&33811&&&33961\tm[4]\\
$6d7p$ &  $^1F^{\rm o}$ &3&34230&155&55&-121&34319& &&34332\tm[4]\\
$7s9s$ & $^3S$             &1&34601&-224&81&-41&34417&34476&-59&\\
\end{tabular}
\end{ruledtabular}
\tablenotetext[1]{NIST data, Ref~\cite{NIST}.}
\tablenotetext[2]{CP+CI, Ref.~\cite{dinh_E120}. IP1 found in
  combination with the calculation for the ion, Ref.~\cite{dinh_E120_ions}.}
\tablenotetext[3]{XIHFSCC, Ref.~\cite{borschevsky}. The value for IP1
  includes a QED radiative and frequency-dependent Breit correction of
  $-46\, {\rm cm}^{-1}$ from Ref. \cite{thierfelder}.}
\tablenotetext[4]{Semi-empirical CP+CI, Ref.~\cite{dzuba_flambaum_Ra}.}
\tablenotetext[5]{IHFSCC,  Ref.~\cite{landau}.}
\end{table*}

The results for radium are presented in Table~\ref{tab:Ra}. It is seen
from the table that there are a number of gaps in the experimental data, and
accurate theoretical predictions of the missing data are important.
We don't agree with all configuration designations used in the
experimental data compiled by NIST \cite{NIST}, and we predict that
the energy of one of the states ($7s8p~^1P_1^{\rm o}$) is
significantly lower than that given in the data tables, as explained later. 

Compared to Ba, the ladder contributions for Ra are smaller for the
terms involving $d$ orbitals in the dominant configurations, while the
contributions to terms from the $ss$ and $sp$ configurations are larger. The Breit contributions 
to the excitation energies range from $10-81\, {\rm cm}^{-1}$ for the levels considered, while 
we saw a significantly larger contribution arising from the QED
radiative corrections, $37-180\, {\rm cm}^{-1}$. 

Unlike the case with Ba, for Ra the term designations change in some
cases when we go from the approximation without ladder diagrams,
CP$^{\rm nl}$+SD+CI, to that with ladder diagrams. 
This means that it is more difficult to track the changes
in energies associated with the ladder corrections, as the 
terms themselves may differ in these approximations.

The deviations from experiment are generally smaller for Ra than for
Ba, with the deviations well under $100\, {\rm cm}^{-1}$ for most
levels; only a few levels deviate more than $100\, {\rm cm}^{-1}$, with the maximum 
deviation $191\, {\rm cm}^{-1}$ for the singlet state $7s6d\, ^1D_2$
(with the exception of $7s8p~{^1P}_1^{\rm o}$, which we address below). 

In the final column of Table~\ref{tab:Ra}, we list the results of other
calculations. For the lower levels, we present the
results of {\it ab initio} calculations, namely from the CP+CI
method \cite{dinh_E120} and the extended intermediate Hamiltonian
Fock-space coupled-cluster (XIHFSCC) method \cite{borschevsky}.   
For the higher levels, where data from these methods is unavailable,
we present the results of the IHFSCC method \cite{landau} and a
semi-empirical CP+CI calculation \cite{dzuba_flambaum_Ra}. 
Both of these calculations miss some of the higher levels that we see. 
Note that QED corrections to the excitation energies were not taken
into account in these other works \cite{dinh_E120,borschevsky,landau}, 
while Breit corrections were also omitted in Ref. \cite{dinh_E120}. 
In Ref. \cite{borschevsky}, the value for IP1 was modified by
adding the frequency-dependent Breit and radiative QED corrections
from Ref. \cite{thierfelder}; the contribution from the radiative
corrections, through use of the radiative potential \cite{radpot}, is $-42\, {\rm
  cm}^{-1}$, in agreement with the current work. 
In Ref. \cite{dzuba_flambaum_Ra}, while Breit and QED radiative
corrections were not included explicitly, these effects were taken
into account to some degree through the empirical fitting factors.

The current calculations for radium are the most complete and most accurate
to date. 
With the high accuracy that we have achieved using this method, we can
be confident of resolving anomalies with experiment in favour of the
theoretical predictions.

The most striking disagreement is with the energy assigned to the
state $7s8p~{^1P}_1^{\rm o}$. In the experimental data \cite{NIST}, 
the large excitation energy $32858\, {\rm cm}^{-1}$ is given to this state. 
However, we do not see this level in our calculations; rather, 
we see a low-lying state, with energy $30703\, {\rm cm}^{-1}$, that is absent in the data.   
In the relativistic regime, the states are defined by their total
angular momentum $J$ and their parity. We see only six levels with 
$J=1$ and odd parity in the energy range we have considered, and 
even extending this range, the next level that we see appears as high as
$36067\, {\rm cm}^{-1}$. 

As for other supporting theoretical calculations for the
${^1P}_1^{\rm o}$ anomaly, results of the semi-empirical CP+CI calculation \cite{dzuba_flambaum_Ra}
included two levels with $J=1$ and odd parity with energies $30695\, {\rm cm}^{-1}$ and 
$31446\, {\rm cm}^{-1}$. These were assigned in that work to the two 
levels $7s8p~^{3}P_1^{\rm o}$ and $7s8p~^{1}P_1^{\rm o}$ in the
experimental data, with very large deviations 
$-868\, {\rm cm}^{-1}$ and $-1412\, {\rm cm}^{-1}$ being seen
there. We suggest that these calculated excitation energies 
should instead be assigned to the low-lying levels $7s8p~{^1P}_1^{\rm o}$ and
$7s8p~{^3P}_1^{\rm o}$. These numbers are in very close agreement with the results of the current
work, $30703\, {\rm cm}^{-1}$ and $31473\, {\rm cm}^{-1}$.
No data for the corresponding singlet state $^1P_1^{\rm o}$ was presented in the
IHFSCC work \cite{landau}. 

Therefore, we suggest that there is no high-lying singlet state
$^1P^{\rm o}_1$ with energy $32858\, {\rm cm}^{-1}$, though we expect
that there should be a lower-lying one with energy around $30700\,
{\rm cm}^{-1}$.

We also suggest that configuration assignments for some terms
be altered. We do not have any terms in our data with leading
configuration $7p^2$ as appears in the experimental data. 
For the triplet terms $7p^2~^3P$, our results indicate instead a
strongly dominating configuration $6d^2~^3P$. 
There is also a case where we see that the dominant configuration is
$7s7d$ rather than $7p^2$, for the term referred in the experimental
data as $7p^2~^1D_2$.

In a system as heavy as radium, and particularly for the
higher levels, the validity of the $LS$ system for designating the
terms loses much of its meaning. Nevertheless, we still believe that
making the previous observations is important, especially when
different assignment of the configurations may lead to confusion
between the designations of levels.

We note further, that with the very strong mixing of configurations in
some terms, these designations become less clear, as we have seen for
the odd-parity $J=1$ terms $7s8p~^1P^{\rm o}_1$, $7s8p~^3P^{\rm o}_1$, and
$6d7p~^3D^{\rm o}_1$, where the mixing between $7s8p$ and $6d7p$ is very strong.

\subsubsection{Element 120}

\begin{table*}
\caption{Ionization potentials (removal of one electron, IP1, and both
  electrons, IP1+IP2) and excitation energies for E120 up to $40000~{\rm cm}^{-1}$. 
CP$^{\rm nl}$+SD+CI excludes ladder diagrams, Breit, and QED; these 
corrections are given in subsequent columns. Results of
other calculations are also presented. Units: ${\rm cm}^{-1}$. Calculated and non-relativistic
g-factors are given in the final columns.}
\label{tab:E120}
\begin{ruledtabular}
\begin{tabular}{llrrrrrrrrrrr}
Conf.& Term & J &CP$^{\rm nl}$&Ladder&Breit& QED & Total &
 CP+CI\tablenotemark[1] & FSCC\tablenotemark[2] &
XIHFSCC\tablenotemark[3] &g&g$_{NR}$\\
 & &&+SD+CI&&&& & && \\
\hline
 IP1 \qquad \ \ $8s^2$&  ${^1S}$ &0&47702&-510&565&-75&47682&47356&47046&47089&& \\
 IP1+IP2 \ $8s^2$&  ${^1S}$ &0& 137807 &-1030&457&-203&137031&137501&136332&136920&& \\
 $8s8p$ & ${^3P}^{\rm o}$ &0&16039&-136&123&-90&15936&15777&15328&15648&0.000& 0.000\\
               &               &1&18035&-148&125&-92&17920&17710&17382&17587&1.409& 1.500\\
                  &                  &2&25924&-338&153&-88&25651&25419&25308&25192&1.499& 1.500\\
 $8s7d$ & ${^3D}$ &1&22904&69&28&-136&22865&22985&22337&22903&0.500& 0.500 \\
                  &              &2&23299&76&42&-134&23283&23163&22494&23034&1.162& 1.167 \\
                  &              &3&23763&96&79&-125&23813&23799&23377&23782&1.333& 1.333 \\
     $8s8p$ & ${^1P}^{\rm o}$ &1&27827&-331&142&-79&27559&27667&28304&27513&1.087& 1.000\\
     $8s7d$ & ${^1D}$ &2&27666&41&95&-133&27669&27438&27652&27247&1.006& 1.000\\
     $8s9s$ & ${^3S}$ &1&31464&-427&145&-63&31119&&&30862&1.998& 2.000\\
   $8s9s$ & ${^1S}$ &0&32633&-431&148&-64&32286&&&&0.000& 0.000\\
     $8s9p$ & ${^3P}^o$  &0&35974&-428&159&-75&35630&&&35463&0.000& 0.000\\
                 &                   &1&36109&-427&159&-77&35764&&&35595&1.369& 1.500\\
                 &                   &2&37908&-462&163&-76&37533&&&37369&1.500& 1.500\\
    $8s8d$  &  ${^3D}$ &1&37880&-427&152&-82&37523&&&&0.501& 0.500 \\
                  &                &2&37932&-426&151&-81&37576&&&&1.145& 1.167\\
                  &              &3&38094&-424&157&-79&37748&&&&1.333& 1.333\\
     $8s9p$  &  $^1P^{\rm o}$ &1&38131&-457&161&-77&37758&&&&1.129& 1.000\\
    $8s8d$ & ${^1D}$ &2&38341&-417&153&-85&37992&&&&1.022& 1.000\\
   $8s10s$ & ${^3S}$ &1&39441&-507&159&-69&39024&&&&1.999& 2.000 \\
   $8p^2/8s10s$ & ${^1S}$ &0&39448&-374&180&-112&39142&&&&0.000& 0.000\\
   $8p^2$/$8s10s$ & ${^1S}$ &0&40137&-341&206&-158&39844&&&&0.000& 0.000\\         
   $8s7f/8s6f$ & ${^3F}^{\rm o}$ &2&40139&-296&167&-104&39906&&&&0.680& 0.667\\
                       &                        &3&40410&-352&168&-77&40149&&&&1.061& 1.083\\ 	
                       &                        &4&40407&-352&168&-76&40147&&&&1.250& 1.250\\	
   $8s7f/8s6f$ & ${^1F}^{\rm o}$ &3&40255&-333&170&-83&40009&&&&1.024&1.000\\
\end{tabular}
\end{ruledtabular}
\tablenotetext[1]{Ref. \cite{dinh_E120}. IP1 found in
  combination with the calculation for the ion, Ref.~\cite{dinh_E120_ions}.}
\tablenotetext[2]{Ref. \cite{skripnikov}.}
\tablenotetext[3]{Ref. \cite{borschevsky}. The value for IP1 includes
  a QED radiative and frequency-dependent Breit correction of $-102\, {\rm cm}^{-1}$ from Ref. \cite{thierfelder}}
\end{table*}

Our results for the ionization potentials and excitation energies for
E120 are presented in Table \ref{tab:E120}. We also present the 
results of our calculations for $g$-factors, and the corresponding
non-relativistic values, to help in identification of the levels. 
For the higher levels, there is strong mixing between configurations,
and we have included the dominant configurations explicitly in the table.

It is seen that the ladder diagrams for E120 are significant, around $400\, {\rm cm}^{-1}$
for many levels, although the maximum
correction for the levels for E120 is less than we saw for the
lighter atoms. The $d$ orbitals are well-screened by the
relativistically contracted $s$ and $p_{1/2}$ orbitals, and there are
no low-lying terms with $d^2$ configuration. 

The Breit corrections to the excitation energies are about $150\, {\rm cm}^{-1}$ for many levels,
while the QED radiative corrections are smaller, reaching $158\, {\rm
  cm}^{-1}$ in the largest case, for the level $8p^2\, ^1S_0$.

In the final columns we list the results of other {\it ab initio} calculations: 
CP+CI \cite{dinh_E120}, a version of FSCC \cite{skripnikov}, and XIHFSCC 
\cite{borschevsky}. We are also aware of another calculation \cite{HDS2013} of the
excitation spectra of E120, though the results are so different from
all other data (by as much as $10000\, {\rm cm}^{-1}$ for some levels) that we have decided
not to present them in the table.

Overall, there is good agreement between the different calculations
presented in the table. It should be noted that Breit and QED corrections
were not included in the calculations \cite{dinh_E120}, while QED was not
included in Refs. \cite{skripnikov,borschevsky}. Also, there is some sensitivity to the
choice of nuclear density. We used the same Fermi distribution that
was used in the previous calculation \cite{dinh_E120}, and in that work the volume
isotopic shifts were studied; we refer the reader to that work for
more details.

The QED radiative correction to the ionization potential IP1 for E120 was calculated recently.
Shabaev {\it et al.} used their model operator approach \cite{shabaev}
to calculate the self-energy correction; they obtained $-202\, {\rm  cm}^{-1}$. 
The total QED correction (including vacuum polarization) was
calculated in Ref. \cite{thierfelder} using the radiative potential
method \cite{radpot}, and they obtained a result of $-77\, {\rm
  cm}^{-1}$, in excellent agreement with our own $-75\, {\rm
  cm}^{-1}$. Their result is broken down into the self-energy and
vacuum polarization contributions, $-183\, {\rm  cm}^{-1}$ and $+106\,
{\rm  cm}^{-1}$, with the former in good agreement with the result of
Shabaev {\it et al}.

The only other calculation of QED corrections to excitation energies
for E120 was performed in the work \cite{HDS2013}. 
The corrections were 
found using the radiative potential method
\cite{radpot} implemented in GRASP \cite{thierfelder,grasp1989,grasp2006}, 
with results in very good agreement with the results of this work.

\subsubsection{Accuracy}

The quality of our calculations can be gauged by comparison of our 
results with experiment. We found excellent agreement between 
theory and experiment for barium and radium, $\lesssim 100\, {\rm cm}^{-1}$ for many 
cases, and we can expect a similar level in the error for our
predictions for E120.

We note that the largest deviations we have seen between theory and 
experiment mostly involve configurations containing $d$-orbitals. 
The smaller deviations we see for Ra compared to Ba could be due to
the $d$-orbitals being screened (also there are fewer configurations
involving $d$-orbitals in the lower levels); this screening effect is even more
pronounced for E120, and therefore the associated errors from
$d$-orbitals may be reduced.

To further study the limits of our method, particularly in relation to
the quality of the correlation potential, 
we have also performed calculations of the spectra for the three atoms
using a different set of screening factors for the Coulomb interaction in the exchange 
part of the correlation potential. The screening factors were  
found from the direct diagrams with both electron-electron screening 
and the hole-particle interaction included, see the second row of Table
\ref{tab:screening}. 
Calculations performed with these modified screening factors gave
results that differed from our final results ``Total'' in
Tables~\ref{tab:Ba}, \ref{tab:Ra}, \ref{tab:E120} over the full
spectral range we considered by about $10-200\, {\rm cm}^{-1}$ for barium, 
about $10-100\, {\rm cm}^{-1}$ for radium, and about $10-200\, {\rm
  cm}^{-1}$ for element 120. 
For E120, the largest deviations occurred for the higher levels
corresponding to configurations comprised of valence 
orbitals that are not the lowest for that wave.
For many of the levels for Ba and Ra,
the results with the adjusted screening factors were better than or as
good as the original results (and often on the other side of
the experimental values). Therefore, we estimate an error that is about half the size
of these shifts, consistent with the error estimates we obtained by looking
at the deviation of the original results from experiment. We expect
similar behaviour for E120.

\section{Conclusion}
 
In summary, we have performed {\it ab initio} calculations of the spectra of Ba, Ra, and E120 using the
recently-developed CP+SD+CI method. We 
have found unprecedented agreement with experiment for Ba and Ra 
and have made accurate predictions for missing and unreliable data and for the spectra of 
E120. 
For Ra, we are confident that the energy
assigned to the level $7s8p\,{^1P}^{\rm o}_1$ in the data
\cite{NIST} is incorrect, and we predict a value $\approx 2200\,
{\rm cm}^{-1}$ smaller. 
The size of the error for our calculations for Ra is about $100\, {\rm cm}^{-1}$ or
better, while it is slightly larger for Ba. We estimate a similar
level of uncertainty for the spectra of E120. 
Finally, we note that the size of the errors in this {\it ab initio}
method is comparable to the size of the QED radiative corrections.

\section*{Acknowledgments}

This work was supported in part by the Australian Research Council.

\end{document}